\documentclass[amssymb,amsmath,fleqn]{iopart}

\usepackage{graphicx}

\def\l{\lambda}
\def\m{\mu}
\def\t{\tau}
\def\d{\delta}
\def\bra{\langle}
\def\ket{\rangle}
\def\a{\alpha}
\def\z{\zeta}

\begin{document}
\title[System-size independence of a LDF for frequency in a 1D
forest-fire model]{System-size independence of a large deviation function for
frequency of events in a one-dimensional forest-fire model with a single
ignition site}

\author{Tetsuya Mitsudo}
\ead{mitsudo@sat.t.u-tokyo.ac.jp} 
\address{FIRST, Aihara Innovative Mathematical Modelling Project,
JST, 4-6-1 Komaba, Meguro, Tokyo, 153-8505, Japan.} 
\address{Institute of Industrial Science, the University of Tokyo,
4-6-1 Komaba, Meguro, Tokyo, 153-8505, Japan.} 
\date{\today}

\begin{abstract}
It is found that a large deviation function for frequency of
 events of size not equal to the system size 
 in the one dimensional forest-fire model with a single ignition site at
 an edge is independent of the system size, by using an exact decomposition
 of the modified transition matrix of a master equation. 
An exchange in the largest eigenvalue of the modified transition matrix
 may not occur in the model.
\end{abstract}

\maketitle

\section{Introduction}

Non-equilibrium phenomena are omnipresent. Sometimes a rare
phenomenon can cause massive effect on our everyday life.
Recently a study on a large deviation
function (LDF), which includes all the higher fluctuations, was
extensively conducted in the context of non-equilibrium statistical
physics \cite{DerL,Ber,Der,Tou,Gia}.
The LDF describes the probability of rare events, and disastrous
events such as large earthquakes can be characterized by the LDF.
Moreover, phase transition between active and inactive phase is
characterized 
by the LDF for an activity in a model of glassy dynamics \cite{Gar}.
The approach using the LDF to dynamic behaviours is soundly progressing
\cite{Gar,Rak,Gie,Coh}, 
however, studies concerning the LDF and criticality are yet developing.

In this study, we apply an approach using the LDF to the dynamic
behaviours of a forest-fire model.
Originally, Bak \etal introduced the forest-fire model to simulate
the system with temporally uniform
injection and fractal dissipation of energy \cite{Bak90}.
The forest-fire model introduced by Bak \etal is later extended by
Drossel and Schwabl \cite{Dro92} in the
context of self organaized criticality (SOC).
Drossel and Schwabl represented forest-fire in the model
with four processes:
planting of trees, ignition, propagation of fire, and extinguishing of
fire.
To separate the timescale of the planting process and those of the
latter three, an effective forest-fire model was introduced to analyze
the model \cite{He93,Pac93}.
The effective model reduces the last three processes to a single
process, the vanishing of a cluster of trees.
Moreover, the effective forest-fire model can be formalized by a master
equation \cite{Hon} and an analytical methods can be applied.

The forest-fire model is also considered as a model of earthquakes
\cite{Tur}.
As an earthquake model, the planting of a tree corresponds to the
loading of stress on the fault and 
 the vanishing of a cluster of trees corresponds
to the triggering of an earthquake which releases the stress on the
connected loaded sites.
A model similar to the one-dimensional forest-fire model with
single triggering site at an edge was introduced as the minimalist model
of earthquakes \cite{Pra}.
The distributions of trigger sites change the
size-frequency distributions of earthquakes in two-dimensional
forest-fire models \cite{Tej}.
Recently, the LDFs for frequency of the largest earthquake in 
forest-fire models with different distributions of trigger
sites was numerically calculated, and a nearly periodic to Poisson
occurrence depending on parameters and distribution of trigger sites is
found \cite{Mits}.

Here, we focus on the model M1, which is the one-dimensional
forest-fire model with single ignition site at an edge \cite{Mits}.
M1 is expressed by a master equation, 
and a standard method to calculate the LDF with
a modified transition matrix of the master equation \cite{Dem} can be
applied to obtain the LDF for frequency of events.
The LDF is given by the Legendre transform of a generating
function which is equal to the largest eigenvalue of the modified
transition matrix.
In this study, we derived an exact decomposition of the modified
transition matrix for the frequency of 
events of size smaller than the system size in M1, by applying a similarity
transformation.
The decomposition implies the system size independence of the LDF for
the frequency.
We numerically calculated the LDF for any system size and compared to that of
the homogeneous Poisson process.
The decomposition enables us to discuss the exchange of the largest and
the second largest eigenvalue of the modified transition matrix in the
limit of infinitely large system size.

We give introduction for the model we use in this study, the LDF and the
method to calculate the LDF. 
Next we derive the decomposition of the modified transition matrix. 
Then, we give numerical calculations of the LDF. In the end, we
present discussions.

\section{Model and Method}

\subsection{Model}

A forest-fire model with single ignition site at an edge 
is called M1 (figure 1). 
M1 can be written by a master equation as,
\begin{equation}
 \frac{{\rm d} P(C;t)}{{\rm d} t} = \sum_{C'}W_L(CC')P(C';t),
\end{equation}
where $C$ denotes the configuration of the system,
$P(C;t)$ is the probability of the system in $C$ at time $t$ and
$W_L(CC')$ is the transition probability rate from $C'$ to $C$ with
system size $L$.
M1 consists of two processes: a loading process and a triggering
process.
The loading process represents the loading of stress onto a fault
and the triggering process represents the occurence of an 
earthquake.
$W_L(CC')$ is denoted $p$ if the transition from $C'$ to $C$ is the
 loading process and is denoted $f$ if the transition is
the triggering process.
We introduce $\t_j\in\{0,1\}$ which represents the state of the site $j$
with $1\leq j\leq L$.
$\t_j=1$ represents loaded state and $\t_j=0$ represents unloaded state.
The configuration $C$ can be written as $C=\{\t_1,\cdots,\t_L\}$.
For example, the system is in the state $C=\{1,1,0,1,0\}$ as in figure 1.
The loading process at the site in the middle is expressed by a
transition from $\{1,1,0,1,0\}$ to $\{1,1,1,1,0\}$. 
The triggering process is expressed by a transition
from $\{1,1,0,1,0\}$ to $\{0,0,0,1,0\}$ and the size of the event is $2$.
A transition from $\{1,1,0,1,0\}$ to $\{1,1,0,0,0\}$ is not allowed in
this model because the triggering only occur from the site $1$.
The transition rates for $C\neq C'$ can be written by $\t_j$
as,
\begin{eqnarray}
\label{wcc}
\fl
& & \fl W_L(CC') = p \sum_{j=1}^L
  [\cdots\d_{\t_{j-1}\t'_{j-1}}(1-\t'_j)\t_j
  \d_{\t_{j+1}\t'_{j+1}}\cdots] \nonumber \\
& & +f\sum_{j=1}^{L} [\t'_1(1-\t_1)\cdots\t'_j(1-\t_j)
   (1-\t'_{j+1})(1-\t_{j+1}) \d_{\t_{j+2}\t'_{j+2}}
   \cdots],
\end{eqnarray}
where $\d_{xx'}$ is the Kronecker delta.
The parts with a suffix less than $1$ or greater than $L$ are omitted.
$W_L(CC')$ for $C=C'$ is written as,
\begin{eqnarray}
\fl & & \fl
 W_L(CC') = -p \sum_{j=1}^L
  [\cdots\d_{\t_{j-1}\t'_{j-1}}(1-\t'_j)(1-\t_j)
  \d_{\t_{j+1}\t'_{j+1}}\cdots] \nonumber \\
& & -f\sum_{j=1}^{L}[\t'_1\t_1\cdots\t'_j\t_j
   (1-\t'_{j+1})(1-\t_{j+1}) \d_{\t_{j+2}\t'_{j+2}}
   \cdots].
\end{eqnarray}
\begin{figure}
\begin{center}
\includegraphics{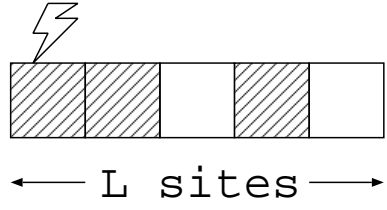}
\caption{A schematic picture of the model M1. Shaded sites represent
 loaded sites, blank sites represent unloaded sites and a thunder mark
 represents an trigger site. If an event is triggered in this
 configuration, the two loaded sites in the left become empty, and the
 size of the event is $2$.}
\end{center}
\end{figure}

The master equation is transformed into a matrix representation as,
\begin{equation}
 \frac{{\rm d}}{{\rm d} t} {\bf P}(t)			
=\mathsf{W}_L {\bf P}(t),
\end{equation}
where ${\bf P}^t(t)=(P_{1}(t),\cdots,P_{2^{L}}(t))$, the index $\m$ of
$P_{\m}(t)$ is given by $\m=\sum_{i=1}^{L}\t_i 2^{i-1}$, 
and $\mathsf{W}_L$ is a transition matrix for the system size $L$.
The index $\m$ has one to one correspondence with the configuration $C$.
For $L=2$, the explicit form of transition matrix is written as,
\begin{equation}
\fl
\mathsf{W}_2=
 \left(\begin{array}{cccc}
-2p & f & 0 & f \\
p & -p-f & 0 & 0 \\
p & 0 & -p & 0 \\
0 & p & p & -f 
\end{array}\right),
\end{equation}
and for $L=3$,
\begin{equation}
\fl
\mathsf{W}_3=
 \left(\begin{array}{cccccccc}
-3p & f & 0 & f & 0 & 0 & 0 & f \\
p & -2p-f & 0 & 0 & 0 & 0 & 0 & 0 \\
p & 0 & -2p & 0 & 0 & 0 & 0 & 0 \\
0 & p & p & -p-f & 0 & 0 & 0 & 0 \\
p & 0 & 0 & 0 & -2p & f & 0 & 0  \\
0 & p & 0 & 0 & p & -p-f & 0 & 0\\
0 & 0 & p & 0 & p & 0 & -p & 0 \\
0 & 0 & 0 & p & 0 & p & p & -f 
\end{array}\right).
\end{equation}

\subsection{Large deviation function}

The mean frequency of events of size $s$ per unit time
$x(s)$ is written as
\begin{equation}
 x(s)=\frac{N(s)}{t}.
\end{equation}
Here, $N(s)$ is the number of events of size $s$ for elapsed
time $t$.
The probability of $x(s)$, $P(x(s))$, asymptotically behaves as
\begin{equation}
 P(x(s)) \sim \exp[-t\phi(x(s))]
\end{equation}
for large $t$, where the function $\phi(x(s))$ is called a large deviation
function (LDF) for the frequency of events of size $s$.
The LDF satisfies $\phi(x_{\rm m}(s))=0$ where $x_{\rm m}(s)$ is the frequency
giving the minimum of $\phi(x(s))$.
A large deviation function $\phi(x(s))$ has a corresponding `generating
function' $\m_s(\l)$, where $\l$ is the conjugate variable of $x(s)$.
The generating function $\m_s(\l)$ is defined as
\begin{equation}
 \exp[\m_s(\l)t]=\bra \exp[\l x(s) t] \ket \sim \int \exp[(\l x'-\phi(x'))t]
  \rmd x'.
\end{equation}
$\phi(x(s))$ and $\m_s(\l)$ are related by the Legendre transform as
\begin{equation}
\label{legendre}
 \phi(x(s))=\max_{\l}[x(s)\l-\m_s(\l)].
\end{equation}
The LDF for the frequency of events in
a homogeneous Poisson process is 
\begin{equation}
\label{ldfpoisson}
 \phi_{\rm P}(x)=x \log(\frac{x}{\a})-x+\a,
\end{equation}
where $\a$ is the rate of event occurrence and the suffix ${\rm P}$ represents
the Poisson process.

To calculate the LDF, the largest eigenvalue of a
modified transition matrix $\mathsf{W}_L^{\l,s}$ is necessary, where
$\l$ is a field related to the number of events of size $s$.
The largest eigenvalue of the modified transition matrix is equal to the
generating function \cite{Dem}.
By using (\ref{legendre}), the LDF is obtained from $\m_s(\l)$.
The modified transition matrix is defined as 
$W^{\l,s}_L(CC')=\exp[\l X(s;CC')]W_L(CC')$, where $X(s;CC')$ is $1$ for the
event of size $s$ and $X(s;CC')=0$ otherwise.
The off-diagonal part of the modified transition rates is written as,
\begin{eqnarray}
\fl
& & \fl W_L^{\l,s}(CC') = p \sum_{j=1}^L
  [\cdots\d_{\t_{j-1}\t'_{j-1}}(1-\t'_j)\t_j
  \d_{\t_{j+1}\t'_{j+1}}\cdots] \nonumber \\
& & \fl +f\sum_{j=1}^{L}[\exp(\l \d_{j,s})\t'_1(1-\t_1)\cdots\t'_j(1-\t_j)
   (1-\t'_{j+1})(1-\t_{j+1}) \d_{\t_{j+2}\t'_{j+2}}
   \cdots] .
\end{eqnarray}
The modified transition matrix for $L=2$ with $s=1$ is written as,
\begin{equation}
\label{mod2}
\fl
\mathsf{W}_2^{\l,1}=
 \left(\begin{array}{cccc}
-2p & f\rme^{\l} & 0 & f \\
p & -p-f & 0 & 0 \\
p & 0 & -p & 0 \\
0 & p & p & -f 
\end{array}\right),
\end{equation}
and for $L=3$,
\begin{equation}
\label{mod3}
\fl
\mathsf{W}_3^{\l,1}=
 \left(\begin{array}{cccccccc}
-3p & f\rme^{\l} & 0 & f & 0 & 0 & 0 & f \\
p & -2p-f & 0 & 0 & 0 & 0 & 0 & 0 \\
p & 0 & -2p & 0 & 0 & 0 & 0 & 0 \\
0 & p & p & -p-f & 0 & 0 & 0 & 0 \\
p & 0 & 0 & 0 & -2p & f\rme^{\l} & 0 & 0 \\
0 & p & 0 & 0 & p & -p-f & 0 & 0\\
0 & 0 & p & 0 & p & 0 & -p & 0 \\
0 & 0 & 0 & p & 0 & p & p & -f 
\end{array}\right).
\end{equation}

\section{System size independence of the LDF}

By observing the forms of $\mathsf{W}_2^{\l,1}$ (\ref{mod2}) and
$\mathsf{W}_3^{\l,1}$ (\ref{mod3}), we find that these two matrices are
related as, 
\begin{equation}
\mathsf{W}_3^{\l,1}=
 \left(\begin{array}{cc}
\mathsf{W}_2^{\l,1}-pI_2 & X_2
\\ pI_2 &
\mathsf{W}_2^{\l,1}-X_2
\end{array}\right),
\end{equation}
where $I_2$ is the identity matrix of size $2^2\times 2^2$ and $X_2$ is
defined as,
\begin{equation}
 X_2=\left(\begin{array}{cccc}
0 & 0 & 0 & f \\
0 & 0 & 0 & 0 \\
0 & 0 & 0 & 0 \\
0 & 0 & 0 & 0 
\end{array}\right).
\end{equation}
By applying a similarity transformation, $\mathsf{W}_3^{\l,1}$ satisfies 
\begin{equation}
\label{eq16}
 \mathsf{W}^{\l,1}_3 \sim 
 \left(\begin{array}{cc}
\mathsf{W}_2^{\l,1}-pI_2-X_2 & -pI_2+X_2
\\ 0 &
\mathsf{W}_2^{\l,1}
\end{array}\right).
\end{equation}
Thus, the eigenvalues of $\mathsf{W}_3^{\l,1}$ are composed of the
eigenvalues of 
$\mathsf{W}_2^{\l,1}$ and $\mathsf{W}_2^{\l,1}-pI_2-X_2$.

Next we derive the relation between $\mathsf{W}^{\l,s}_{L+1}$ and
$\mathsf{W}^{\l,s}_{L}$.
The $\mathsf{W}^{\l,s}_{L+1}$ is written as,
\begin{eqnarray}
\fl
& & \fl W_{L+1}^{\l,s}(CC') = p \sum_{j=1}^{L+1}
  [\cdots\d_{\t_{j-1}\t'_{j-1}}(1-\t'_j)\t_j
  \d_{\t_{j+1}\t'_{j+1}}\cdots] \nonumber \\
& & \fl +f\sum_{j=1}^{L+1}[\exp(\l\d_{j,s})\t'_1(1-\t_1)\cdots\t'_j(1-\t_j)
   (1-\t'_{j+1})(1-\t_{j+1}) \d_{\t_{j+2}\t'_{j+2}}
   \cdots] \nonumber \\
& & \fl -p \sum_{j=1}^{L+1}
  [\cdots\d_{\t_{j-1}\t'_{j-1}}(1-\t'_j)(1-\t_j)
  \d_{\t_{j+1}\t'_{j+1}}\cdots] \nonumber \\
& & \fl -f\sum_{j=1}^{L+1}[\t'_1\t_1\cdots\t'_j\t_j
   (1-\t'_{j+1})(1-\t_{j+1}) \d_{\t_{j+2}\t'_{j+2}}
   \cdots].
\end{eqnarray}
$\mathsf{W}^{\l,s}_{L+1}$ is written by using $\mathsf{W}^{\l,s}_{L}$
as,
\begin{eqnarray}
 \mathsf{W}^{\l,s}_{L+1} &=&
\mathsf{W}^{\l,s}_L\d_{\t_{L+1}\t'_{L+1}}+
p \d_{\t_1\t'_1}\cdots\d_{\t_{L}\t'_{L}} (1-\t'_{L+1})\t_{L+1}
\nonumber \\
& &
-p \d_{\t_1\t'_1}\cdots\d_{\t_{L}\t'_{L}}(1-\t'_{L+1})(1-\t_{L+1})
\nonumber \\
& & +f \t'_1(1-\t_1)\cdots\t'_L(1-\t_L)
   \t'_{L+1}(1-\t_{j+1}) \nonumber \\
& & -f \t'_1(1-\t_1)\cdots\t'_L(1-\t_L)\t_{L+1}\t'_{L+1}
\label{eq18}
\end{eqnarray}
(\ref{eq18}) is also written in the matrix form as,
\begin{eqnarray}
\label{recrel}
\fl & & \fl
  \mathsf{W}^{\l,s}_{L+1} = \left(\begin{array}{cc}
\mathsf{W}^{\l,s}_L -p \d_{\t_1\t'_1}\cdots\d_{\t_L\t'_L} &
f \t'_1(1-\t_1)\cdots\t'_L(1-\t_L) \\
p \d_{\t_1\t'_1}\cdots\d_{\t_L\t'_L} &
\mathsf{W}^{\l,s}_L - f \t'_1(1-\t_1)\cdots\t'_L(1-\t_L)
\end{array}\right) \\
&=& 
\label{eqrel}
\left(\begin{array}{cc}
\mathsf{W}^{\l,s}_L -p I_L & X_L \\
p I_L &
\mathsf{W}^{\l,s}_L - X_L
\end{array}\right).
\end{eqnarray}
where $I_L=\prod_{i=1}^L\d_{\t_{i}\t'_{i}}$ and 
\begin{equation}
 X_L=\left(\begin{array}{ccc}
0 & \cdots & f \\
\vdots & \ddots & \vdots \\
0 & \cdots & 0 
\end{array}\right).
\end{equation}
$X_L$ corresponds to the term $f\t'_1(1-\t_1)\cdots\t'_L(1-\t_L)$.
A similarity transformation leads to 
\begin{equation}
\label{recrel}
 \mathsf{W}^{\l,s}_{L+1} \sim 
 \left(\begin{array}{cc}
\mathsf{W}_L^{\l,s}-pI_L-X_L
 & -pI_L+X_L
\\ 0 &
\mathsf{W}_L^{\l,s}
\end{array}\right).
\end{equation}
Thus, the eigenvalues of $\mathsf{W}_{L+1}^{\l,s}$ are composed of the
eigenvalues of 
$\mathsf{W}_L^{\l,s}$ and $\mathsf{W}_L^{\l,s}-pI_L-X_L$.

From (\ref{eqrel}), $\mathsf{W}_L^{\l,s}-pI_L-X_L$ is written as,
\begin{equation}
\label{decomp}
\fl
 \mathsf{W}_L^{\l,s}-pI_L-X_L=
\left(\begin{array}{cc}
\mathsf{W}_{L-1}^{\l,s}-2pI_{L-1} & 0 \\
pI_{L-1} & \mathsf{W}_{L-1}^{\l,s}-pI_{L-1}-X_{L-1}
\end{array}\right).
\end{equation}
By using (\ref{recrel}) and
(\ref{decomp}), $\mathsf{W}_L^{\l,s}$ is
decomposed into $\mathsf{W}_{s+1}^{\l,s}-kpI_{s+1}$ for even $k$ and
$\mathsf{W}_{s+1}^{\l,s}-kpI_{s+1}-X_{s+1}$ for odd $k$ with
$k=\{0,\cdots,L-s-1\}$,
where each component is degenerate $_{L-s-1}C_k$ times.
Here we mean by decomposition that a set of eigenvalues of a matrix is
decomposed into sets of eigenvalues of the component matrices.
For example, from (\ref{eq16}), $\mathsf{W}_3^{\l,1}$ is decomposed into 
$\mathsf{W}_2^{\l,1}$ and $\mathsf{W}_2^{\l,1}-pI_2-X_2$.
The proof of the decomposition of $\mathsf{W}_L^{\l,s}$ is given in the
appendix.

The decomposition of $\mathsf{W}_L^{\l,s}$ suggests that the largest
eigenvalue of $\mathsf{W}_L^{\l,s}$ is included in the eigenvalues of
$\mathsf{W}_{s+1}^{\l,s}$ or $\mathsf{W}_{s+1}^{\l,s}-pI_{s+1}-X_{s+1}$
for any $L>s$.
The largest eigenvalues of the other components are smaller than the two
components, because the term $-pkI_L$ just shifts all the eigenvalues.
Among the decomposed elements,
the eigenvalues of $\mathsf{W}_{2}^{\l,1}-pI_{2}-X_{2}$, which denote
$\z_1,\cdots,\z_4$, are calculated as,
\begin{eqnarray}
\z_1 &=&-2p \\
\z_2 &=&-p-f \\
\z_3 &=& \frac{-5p-f+\sqrt{(p-f)^2+4pf\rme^{\l}}}{2} \\
\z_4 &=& \frac{-5p-f-\sqrt{(p-f)^2+4pf\rme^{\l}}}{2} .
\end{eqnarray}
However, the analytical forms of the eigenvalues of other cases
are complex.
For small $s$ and $L$, we can numerically calculate the
largest eigenvalue of $\mathsf{W}_{s+1}^{\l,s}$ or
$\mathsf{W}_{s+1}^{\l,s}-pI_{s+1}-X_{s+1}$.

The numerical calculations suggest that the largest eigenvalue of
 $\mathsf{W}_2^{\l,1}$ is larger than that of
 $\mathsf{W}_2^{\l,1}-pI_2-X_2$ (figure 2), for $0.01\leq f\leq 1$ and
 $-3\leq\l\leq 5$.
Thus, the largest eigenvalue of $\mathsf{W}_2^{\l,1}$ is the largest
 eigenvalue of $\mathsf{W}_L^{\l,1}$.
We assume that this holds for other $f$ and $\l$.
\begin{figure}
\begin{center}
\includegraphics[scale=0.5]{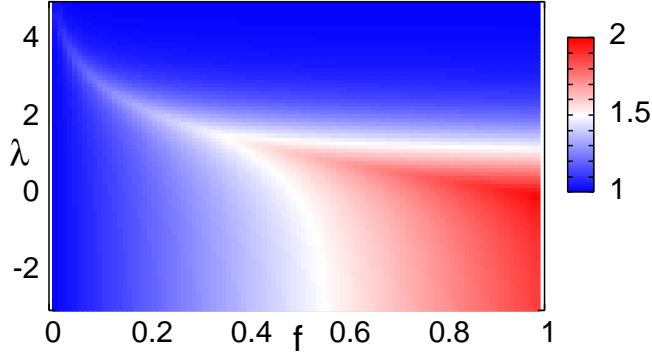}
\caption{The difference between the largest eigenvalue of
 $\mathsf{W}_2^{\l,1}$ and that of $\mathsf{W}_2^{\l,1}-pI_2-X_2$ for 
$0.01\leq f\leq 1$ and $-3\leq\l\leq 5$. The difference is higher than
 $1$ for this parameter region which suggest the largest eigenvalue of
 $\mathsf{W}_2^{\l,1}$ is larger than that of
 $\mathsf{W}_2^{\l,1}-pI_2-X_2$.}
\end{center}
\end{figure}

The generating function corresponding to the LDF for frequency of
events of size $s$ is equal to the largest eigenvalue of
$\mathsf{W}^{\l,s}_L$.
In figure 3(a), $\phi(z(1))/x_{\rm m}(1)$ for $L=2,6,10$ is
plotted with $f=1.0,0.1$ and $0.01$. 
$z(s)$ is defined as $z(s)=(x(s)-x_{\rm m}(s))/x_{\rm m}(s)$.
The numerically calculated LDFs are exactly the same as that of
different $L$.
The solid line denotes the LDF for frequency of a homogeneous Poisson
process in $z$, $\phi_{\rm P}(z)=(z+1)\log(z+1)-z$.
The LDFs for $f=0.1$ and $0.01$ are below $\phi_{\rm P}(z)$, 
which suggests that the fluctuation is larger than that of the Poisson
process.
\begin{figure}
\begin{center}
\includegraphics[scale=0.4]{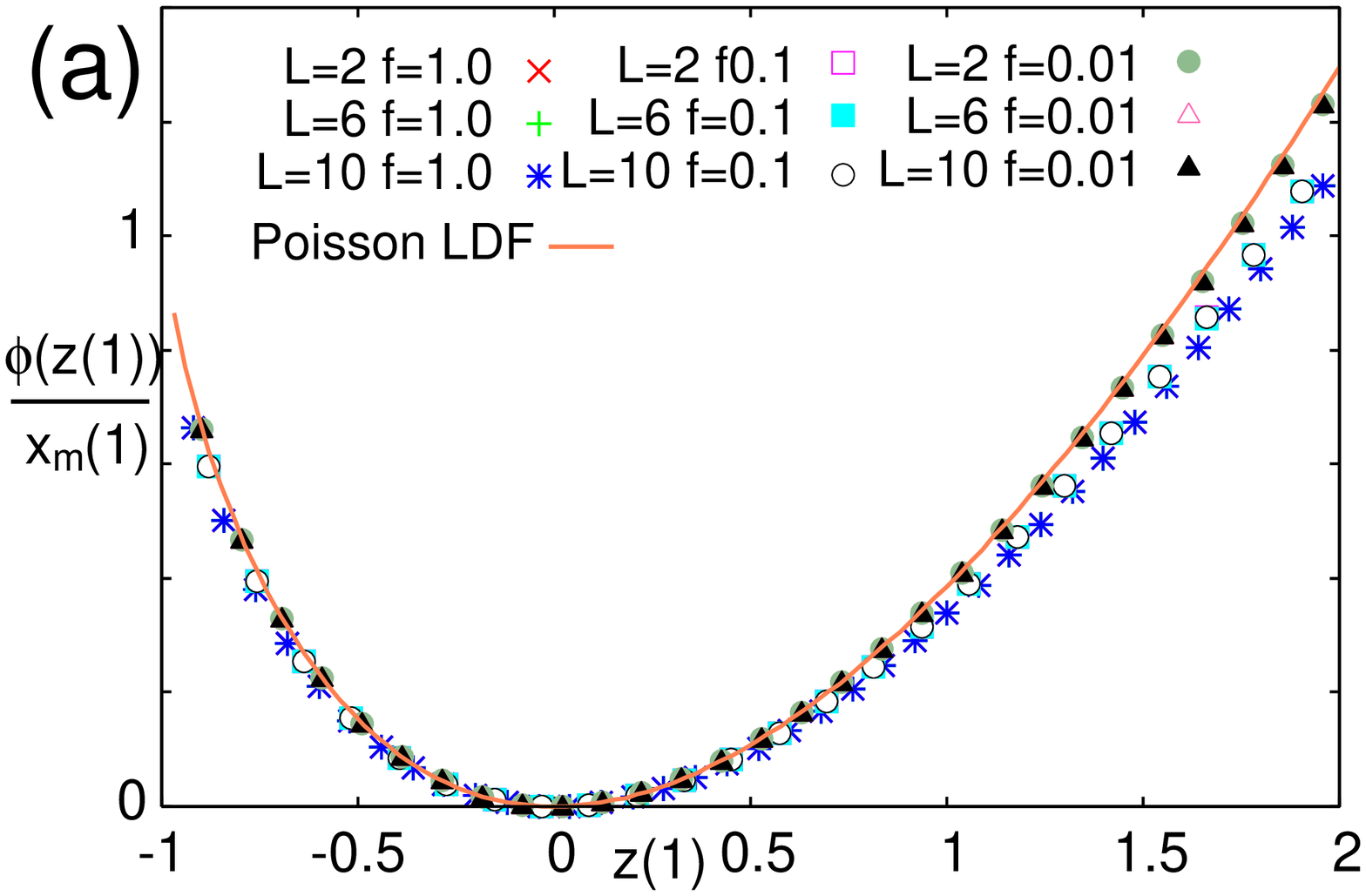}
\includegraphics[scale=0.4]{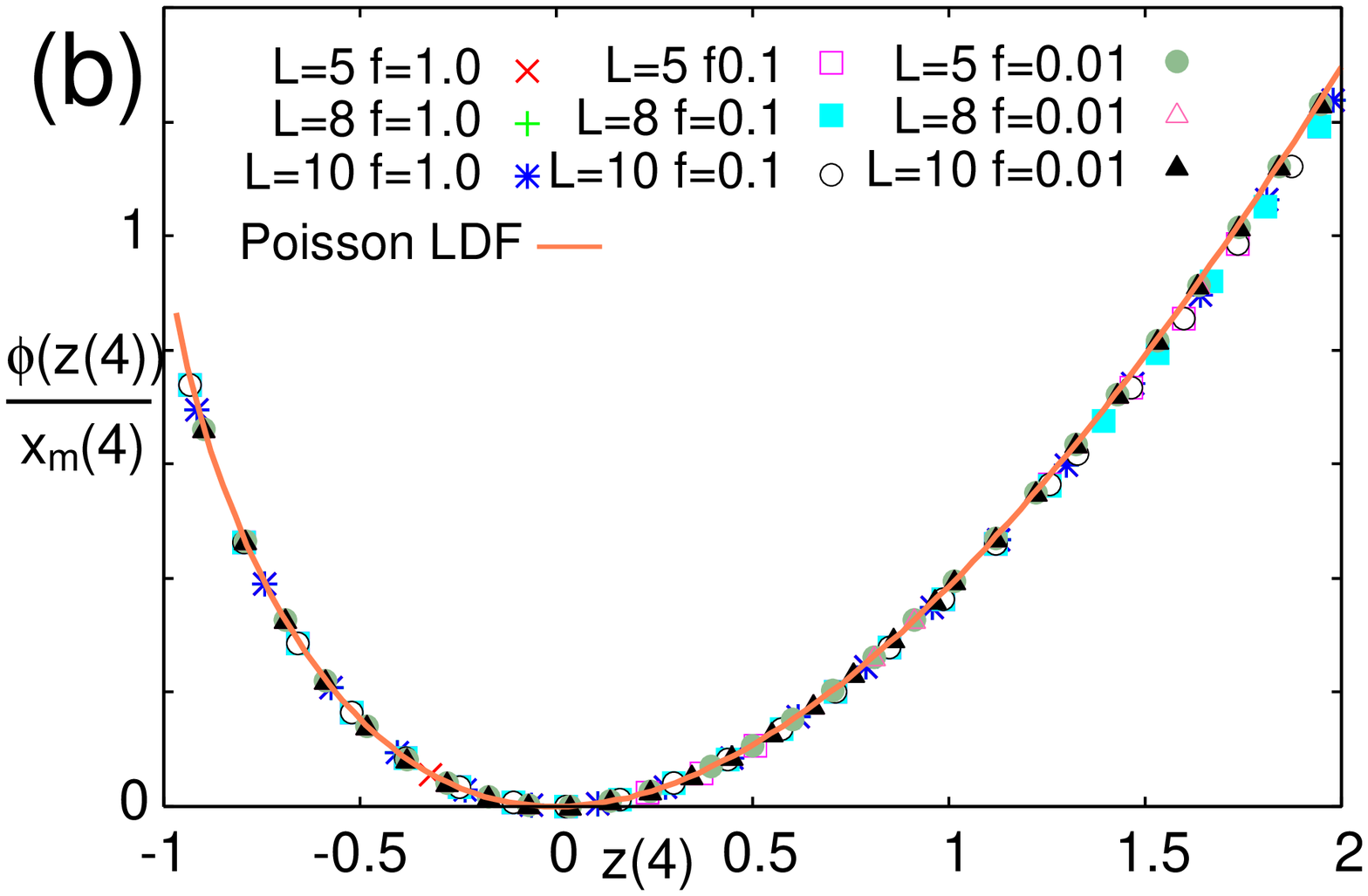}
\caption{(a) $\phi(z(1))/x_{\rm m}(1)$ for $L=2,6,10$ and (b)
 $\phi(z(4))/x_{\rm m}(4)$ for $L=5,8,10$ with $f=1.0,0.1$ and
 $0.01$. The LDF  takes the same value for different $L$.}
\end{center}
\end{figure}
In the figure 3(b), the LDF for frequency of $s=4$ is
presented for $L=5,8,10$ and $f=1.0,0.1$ and $0.01$.
The LDFs in the figure 3(b) is well approximated by the Poisson LDF.

\section{Discussions}

The decomposition of $\mathsf{W}_{L}^{\l,s}$ holds for any $L>s$.
If we increase $L$ and continue the decomposition for many times, a
newly produced component will have a lower shifted term, so that the
eigenvalues are always smaller than that of $\mathsf{W}_{s+1}^{\l,s}$ or
$\mathsf{W}_{s+1}^{\l,s}-pI_{s+1}-X_{s+1}$.
Thus, the LDF or the largest
eigenvalue of the modified transition matrix in the 
$L\to\infty$ limit is still the largest eigenvalue of
$\mathsf{W}_{s+1}^{\l,s}$ or $\mathsf{W}_{s+1}^{\l,s}-pI_{s+1}-X_{s+1}$.
For $s=1$, the claim that the largest eigenvalue is included in
$\mathsf{W}_{2}^{\l,1}$ is supported by the numerical calculations.
It is an open problem whether the largest eigenvalue is always included in
$\mathsf{W}_{s+1}^{\l,s}$ or not.
If the largest eigenvalue is exchanged with the other eigenvalues in the
$L\to\infty$ limit, a dynamical phase transition may occur.
A criticality for $L\to\infty$ is an important
problem in the self organized criticality (SOC) point of
view, since there has been intensive studies concerning the SOC of the
forest-fire models based on simulations, for example see \cite{Gra,Bon}.
Note that M1 is different from usual forest-fires in the distribution of
trigger sites and in the absence of fire propagation process.
It is an interesting development that if the approach based on the LDF
in this study can contribute to such exciting problems.
The decomposition found in this study is limited to M1, however the same
structure might be found in other models written by master equations.
The exchange of the largest eigenvalue and the second largest eigenvalue
can be also discussed by numerical calculations of the models of
interest, so that future development concerning the LDF and related
phase transitions is largely expected.

We found an exact decomposition of the modified transition matrix
concerning the frequency of events of size $s<L$ in the model M1.
The decomposition leads to the independence of $L$ in
the LDF for frequency of events of size $s<L$.
However, for the events of size $L$, 
the LDF for frequency depend on $L$ \cite{Mits}.
Concerning earthquakes, relation between the LDF of
frequency of small earthquakes and that of the system-size earthquakes
is an interesting problem because the relation between them is related
to the problem of capturing the symptoms of large earthquakes.

\ack

The author would like to thank M. Katori for stimulating comments, and
N. Kato for fruitful discussions.
This work is supported by the Aihara Project, the FIRST program from the
 JSPS, initiated by the CSTP.

\appendix

\section{Proof of the decomposition}

We derived a similarity relation
\begin{equation}
\label{recrela}
\fl
 \mathsf{W}^{\l,s}_{L} \sim 
 \left(\begin{array}{cc}
\mathsf{W}_{L-1}^{\l,s}-pI_{L-1}-X_{L-1}
 & -pI_{L-1}+X_{L-1}
\\ 0 &
\mathsf{W}_{L-1}^{\l,s}
\end{array}\right)
\end{equation}
 and an equation 
\begin{equation}
\label{decompa}
\fl
 \mathsf{W}_L^{\l,s}-pI_L-X_L=
\left(\begin{array}{cc}
\mathsf{W}_{L-1}^{\l,s}-2pI_{L-1} & 0 \\
pI_{L-1} & \mathsf{W}_{L-1}^{\l,s}-pI_{L-1}-X_{L-1}
\end{array}\right)
\end{equation}
for $\mathsf{W}_L^{\l,s}$.
Using these two relations, we prove a proposition that
'$\mathsf{W}_L^{\l,s}$ is decomposed into $\mathsf{W}_{s+1}^{\l,s}-kpI_{s+1}$
for even $k$ and $\mathsf{W}_{s+1}^{\l,s}-kpI_{s+1}-X_{s+1}$ for odd $k$
which are degenerate $_{L-s-1}C_k$ times where $k=\{0,\cdots,L-s-1\}$' by
the mathematical induction.

For $L=s+2$, $\mathsf{W}^{\l,s}_{s+2}$ is decomposed into
$\mathsf{W}^{\l,s}_{s+1}$ and $\mathsf{W}^{\l,s}_{s+1}-pI_{s+1}-X_{s+1}$.
For $L=s+3$, $\mathsf{W}^{\l,s}_{s+3}$ is decomposed into
$\mathsf{W}^{\l,s}_{s+1}$, two of
$\mathsf{W}^{\l,s}_{s+1}-pI_{s+1}-X_{s+1}$ an
$\mathsf{W}^{\l,s}_{s+1}-2pI_{s+1}$.
For $L=s+4$, from (\ref{recrela}) and (\ref{decompa}), 
\begin{equation}
\fl
 \mathsf{W}_{s+4}^{\l,s} \sim
\left(\begin{array}{cc}
\mathsf{W}_{s+3}^{\l,s}-pI_{s+3}-X_{s+3}
 & -pI_{s+3}+X_{s+3}
\\ 0 &
\mathsf{W}_{s+3}^{\l,s}
\end{array}
\right)
\end{equation}
and
\begin{equation}
\fl
 \mathsf{W}_{s+3}^{\l,s}-pI_{s+3}-X_{s+3}=
\left(\begin{array}{cc}
\mathsf{W}_{s+2}^{\l,s}-2pI_{s+2} & 0 \\
pI_{s+2} & \mathsf{W}_{s+2}^{\l,s}-pI_{s+2}-X_{s+2}
\end{array}\right),
\end{equation}
are satisfied.
Thus, $\mathsf{W}_{s+4}^{\l,s}$ is decomposed into
$\mathsf{W}_{s+3}^{\l,s},\mathsf{W}_{s+2}^{\l,s}-2pI_{s+2}$ and 
$\mathsf{W}_{s+2}^{\l,s}-pI_{s+2}-X_{s+2}$.
$\mathsf{W}_{s+2}^{\l,s}-2pI_{s+2}$ is decomposed into
$\mathsf{W}^{\l,s}_{s+1}-3pI_{s+1}-X_{s+1}$ and
$\mathsf{W}^{\l,s}_{s+1}-2pI_{s+1}$.
$\mathsf{W}_{s+2}^{\l,s}-pI_{s+2}-X_{s+2}$ is decomposed into
$\mathsf{W}_{s+1}^{\l,s}-pI_{s+1}-X_{s+1}$ and
$\mathsf{W}_{s+1}^{\l,s}-2pI_{s+1}$.
By collecting all the components, $\mathsf{W}_{s+4}^{\l,s}$ is decomposed into
$\mathsf{W}_{s+1}^{\l,s}$, $3$ of
$\mathsf{W}_{s+1}^{\l,s}-pI_{s+1}-X_{s+1}$, $3$ of
$\mathsf{W}_{s+1}^{\l,s}-2pI_{s+1}$ and
$\mathsf{W}_{s+1}^{\l,s}-3pI_{s+1}-X_{s+1}$ (See table A1.).
Thus, the proposition holds for $L=s+4$.

\begin{table}[htb]
\caption{Decomposition table for $L=3,\cdots,6$ and for odd
 $L$. $\mathsf{W}_L^{\l,s}$ is decomposed into
 the matrices in the left column, and the number of degeneracy is
 presented.}
\begin{indented}
\item[]\fl\begin{tabular}{c|cccccc} \br
 & $\mathsf{W}_{s+2}^{\l,s}$ & $\mathsf{W}_{s+3}^{\l,s}$ &
 $\mathsf{W}_{s+4}^{\l,s}$
 & $\mathsf{W}_{s+5}^{\l,s}$ & $\cdots$ & $\mathsf{W}_L^{\l,s}$ \\ \mr
$\mathsf{W}_{s+1}^{\l,s}$ & 1 & 1 & 1 & 1 & $\cdots$ & ${}_{L-s-1}C_0$ \\ 
$\mathsf{W}_{s+1}^{\l,s}-pI_{s+1}-X_{s+1}$ & 1 & 2 & 3 & 4 & $\cdots$ &
			 ${}_{L-s-1}C_1$ \\ 
$\mathsf{W}_{s+1}^{\l,s}-2pI_{s+1}$ & 0 & 1 & 3 & 6 & $\cdots$ & ${}_{L-s-1}C_2$ \\ 
$\mathsf{W}_{s+1}^{\l,s}-3pI_{s+1}-X_{s+1}$ & 0 & 0 & 1 & 4 & $\cdots$ &
			 ${}_{L-s-1}C_3$ \\ 
$\vdots$ & $\vdots$ & $\vdots$ & $\vdots$ & $\vdots$ & $\ldots$ &
			 $\vdots$ \\ 
$\mathsf{W}_{s+1}^{\l,s}-(L-s-2)pI_{s+1}$ & 0 & 0 & 0 & 0 & $\cdots$ &
			 ${}_{L-s-1}C_{L-s-2}$ \\ 
$\mathsf{W}_{s+1}^{\l,s}-(L-s-2)pI_{s+1}-X_{s+1}$ & 0 & 0 & 0 & 0 & $\cdots$ &
			 ${}_{L-s-1}C_{L-s-1}$ \\ \br
\end{tabular}
\end{indented}
\end{table}

Let us assume that $\mathsf{W}_L^{\l,s}$ is decomposed into 
$\mathsf{W}_{s+1}^{\l,s}-kpI_{s+1}$ for even $k$ and 
$\mathsf{W}_{s+1}^{\l,s}-kpI_{s+1}-X_{s+1}$ for odd $k$ which are degenerate
${}_{L-s-1}C_k$ times where $k=\{0,\cdots,L-s-1\}$,
and also $\mathsf{W}_{L-1}^{\l,s}$
is decomposed into $\mathsf{W}_{s+1}^{\l,s}-kpI_{s+1}$ for even $k$ and 
$\mathsf{W}_{s+1}^{\l,s}-kpI_{s+1}-X_{s+1}$ for odd $k$ which are degenerate
${}_{L-s-2}C_k$ times where $k=\{0,\cdots,L-s-2\}$.
$\mathsf{W}_{L+1}^{\l,s}$ is composed of
$\mathsf{W}_{L}^{\l,s}-pI_L-X_L$ and $\mathsf{W}_{L}^{\l,s}$.
$\mathsf{W}_{L}^{\l,s}-pI_L-X_L$ is composed of
$\mathsf{W}_{L-1}^{\l,s}-pI_{L-1}-X_{L-1}$ and
$\mathsf{W}_{L-1}^{\l,s}-2pI_{L-1}$.
By subtracting the components of $\mathsf{W}_{L-1}^{\l,s}-2pI_{L-1}$ from
the components of $\mathsf{W}_L^{\l,s}$,
it directly follows that
$\mathsf{W}^{\l,s}_{L-1}-pI_{L-1}-X_{L-1}$ is decomposed into
$\mathsf{W}_{s+1}^{\l,s}-kpI_{s+1}$ for even $k$ and 
$\mathsf{W}_{s+1}^{\l,s}-kpI_{s+1}-X_{s+1}$ for odd $k$ which are degenerate
${}_{L-3}C_{k-1}$ times where $k=\{1,\cdots,L-s-1\}$.
Also using the assumption, 
$\mathsf{W}_{L-1}^{\l,s}-2pI_{L-1}$ is decomposed into
$\mathsf{W}_{s+1}^{\l,s}-(k+2)pI_{s+1}$ for even $k$ and
$\mathsf{W}_{s+1}^{\l,s}-(k+2)pI_{s+1}-X_{s+1}$ for odd $k$ 
which are degenerate ${}_{L-3}C_{k-2}$ times where $k=\{2,\cdots,L-s\}$.
The number of degeneracy and the decomposed elements are summarized in the
table A2.
\begin{table}[htb]
\caption{The decomposition table for calculating
 $\mathsf{W}_{L+1}^{\l,s}$. The table is for odd $L$.
}
\begin{indented}
\item[]\fl\begin{tabular}{@{}c|@{}c@{}c@{}c@{}c} \br
 component & $\mathsf{W}_L^{\l,s}$ &
 $\mathsf{W}_{L-1}^{\l,s}-pI_{L-1}-X_{L-1}$ &
 $\mathsf{W}_{L-1}^{\l,s}-2pI_{L-1}$ & $\mathsf{W}_{L+1}^{\l,s}$ \\ \mr
 $\mathsf{W}_{s+1}^{\l,s}$ & ${}_{L-s-1}C_0$ & 0  & 0 & ${}_{L-s}C_0$ \\ 
 $\mathsf{W}_{s+1}^{\l,s}-pI_{s+1}-X_{s+1}$ & ${}_{L-s-1}C_1$  &
	 ${}_{L-s-2}C_0$  & 0 & ${}_{L-s}C_1$ \\ 
 $\mathsf{W}_{s+1}^{\l,s}-2pI_{s+1}$ & ${}_{L-s-1}C_2$ & ${}_{L-s-2}C_1$ &
	     ${}_{L-s-2}C_0$ & ${}_{L-s}C_2$ \\ 
 $\mathsf{W}_{s+1}^{\l,s}-3pI_{s+1}-X_{s+1}$ & ${}_{L-s-1}C_3$  &
	 ${}_{L-s-2}C_2$  & ${}_{L-s-2}C_1$ & ${}_{L-s}C_3$ \\ 
 $\vdots$ & $\vdots$ & $\vdots$ & $\vdots$ & $\vdots$ \\ 
 $\mathsf{W}_{s+1}^{\l,s}-kpI_{s+1}$ & ${}_{L-s-1}C_{k}$  & ${}_{L-s-2}C_{k-1}$
	     & ${}_{L-s-2}C_{k-2}$ & ${}_{L-s}C_k$ \\ 

 $\mathsf{W}_{s+1}^{\l,s}-(k+1)pI_{s+1}-X_{s+1}$ & ${}_{L-s-1}C_{k+1}$ &
	 ${}_{L-s-2}C_{k}$  & ${}_{L-s-2}C_{k-1}$ & ${}_{L-s}C_{k+1}$ \\ 
 $\vdots$ & $\vdots$ & $\vdots$ & $\vdots$ & $\vdots$ \\ 
 $\mathsf{W}_{s+1}^{\l,s}-(L-s-2)pI_{s+1}$ & ${}_{L-s-1}C_{L-s-2}$ &
	     ${}_{L-s-2}C_{L-s-3}$ & ${}_{L-s-2}C_{L-s-4}$ &
		 ${}_{L-s}C_{L-s-2}$ \\ 
 $\mathsf{W}_{s+1}^{\l,s}-(L-s-1)pI_{s+1}-X_{s+1}$ & ${}_{L-s-1}C_{L-s-1}$ &
	 ${}_{L-s-2}C_{L-s-2}$  & ${}_{L-s-2}C_{L-s-3}$ &
		 ${}_{L-s}C_{L-s-1}$ \\ 
 $\mathsf{W}_{s+1}^{\l,s}-(L-s)pI_{s+1}$ & $0$  & $0$  & ${}_{L-s-2}C_{L-s-2}$
		 & ${}_{L-s}C_{L-s}$ 
		 \\ \br
\end{tabular}
\end{indented}
\end{table}
As we see in the table A2, we can sum up the number of degeneracy of
$\mathsf{W}_{s+1}^{\l,s}-kpI_{s+1}$ as
${}_{L-s-1}C_{k}+{}_{L-s-2}C_{k-1}+{}_{L-s-2}C_{k-2}$.
The summation is equal to ${}_{L-s}C_k$ which is the same as the number
of degeneracy for $\mathsf{W}_{L+1}^{\l,s}$.
Thus, the decomposition of the $L+1$ matrix satisfy the proposition.

\end{document}